\documentclass[12pt,manuscript]{aastex}
\begin{document}
\title {Reverberation in the UV-Optical Continuum Brightness 
Fluctuations of MACHO Quasar 13.5962.237}
\author{ Rudolph E. Schild\footnote{Harvard-Smithsonian Center for Astrophysics, 60 Garden
Street,
Cambridge, MA 02138, USA: rschild@cfa.harvard.edu; corresponding author,
phone 617-495-7426: FAX 617-495-7467}, Justin Lovegrove
  \footnote{Harvard-Smithsonian Center for
  Astrophysics and Southampton University} \& Pavlos 
  Protopapas\footnote{Center for Astrophysics and
  Department of Physics, Harvard University}}
\begin{abstract}
 We examine the nature of brightness fluctuations in the
UV-Optical spectral region of an ordinary quasar with 881 optical
brightness measurements made during the epoch 1993 - 1999. We find evidence
for systematic trends having the character of a pattern of reverberations
following an initial disturbance. The initial pulses have brightness
increases of order 20\% and pulse widths of 50 days, and the reverberations 
have typical amplitudes of 12\% with longer mean pulse widths of order
80 days and pulse separations of order 90 days.
The repeat pattern occurs over the same
time scales whether the initial disturbance is a brightening or fading.
The lags of the pulse trains are comparable to the lags seen previously in
reverberation of the broad blue-shifted emission lines following brightness
disturbances in Seyfert galaxies, when allowance is made for the mass of
the central object. In addition to the burst pulse trains, we find evidence
for a semi-periodicity with a time scale of 2 years. These strong patterns
of brightness fluctuations suggest a method of discovering quasars
from photometric monitoring alone, with data of the quality expected from
large brightness monitoring programs like Pann-Stars and LSST.

\end{abstract}
\section{Introduction}

Shortly after the discovery of quasars, brightness fluctuations of their
UV-optical brightnesses were discovered (Matthews and Sandage, 1963).
But while several important brightness monitoring
programs were undertaken to observe such fluctuations with photographic
and photoelectric photometry, little is yet known about the nature and 
origin of such fluctuations. Many studies searching for short term
inter-night and intra-night variability have been undertaken, especially
for BL Lac objects, to put constraints on internal structure, possibly the
inner edge of the accretion disc. We concern ourselves here with longer
term variability because it would evidence the global structure of the
quasar. 

Among the first and most comprehensive studies was the publication by
Hawkins (1996), based upon a sparse sampling 
(4 observations per year). A more recent work (Hawkins 2007)
incorporated an analysis of
the MACHO quasars with structure function analysis and 
Fourier techniques not suited for the study of
random events of the kind we find. Netzer et al (1996)
analyzed photoelectric photometry and looked for many correlations with
quasar properties but did not find the reverberation signature.
Webb and Malkan (2000) analyzed brightness fluctuations but
on time scales too short to recognize the reverberation patterns that we
find in our data and Rengstorf et al (2004) used structure function 
analysis for detection of variability in large samples of stellar objects. 

The existence of reverberation in the UV-optical continuum was recognized
already in 1997, and interpreted as indicative of large quasar
structure by Schild, Leiter, and Robertson (2006) in the comprehensively
studied gravitationally lensed Q0957 (The First Lens) quasar. Following the
earlier discovery by Thomson and Schild (1997) from autocorrelation
analysis that strong repeat patterns of brightness fluctuations were seen
over a long 20-year period of monitoring, these reverberations were
combined with microlensing results to produce a picture of luminous
structure that seemed to reinforce the picture of outer quasar structure 
already found by Elvis (2000). In particular, the radial distance of the
Elvis outflow structure matched the distance to the reverberating emission
line structures found in seyfert galaxy NGC 5548 (Peterson et al 1993) when
allowance was made for the larger quasar mass.
A refinement by Schild (2005) seemed to
demonstrate that the pulse trains observed evidenced outer structure above
and below the accretion disc plane, about as described in the Elvis (2000)
model. A microlensing simulation showing that the inner and outer luminous
structure is required to fit all available microlensing data was given by
Schild \& Vakulik (2003).

No physical theory is yet available to guide a discussion of such large
quasar structure as implied if the reverberations, observed on time scales
of approximately 100 days, are produced by the collapsed central
object and propagate to outer luminous structure at light speed. 
The broad, blue-shifted high-excitation emission lines present in
all quasars are presumed in standard theory to originate in clouds 
randomly orbiting the central
structure. A unification model featuring absorption in a dusty torus has no
physical basis in kinematic or dynamical theory, although it seems to be
required to explain the outflow winds that are revealed in diverse 
spectroscopic data
(Elvis, 2000). Proga (2000) has developed a theory of line
driving to explain the outward forces driving an outflow wind but no
theory is available to explain why mass is observed so high above the
accretion disc plane. However the MECO model of black hole and quasar 
structure seems to offer a way for strong magnetic fields originating at
the center to produce the uplift as a result of magnetic effects caused
near the outer light cylinder (Schild, Leiter, and Robertson 2008). In this
case the central object would be radiatively inefficient (Robertson and
Leiter, 2006), much as standard black hole models also predict.

\section{ Data and Preliminary Processing} 

We have analyzed the data for the 59 quasars that are seen behind the
Magellanic Clouds and which therefore had their brightnesses monitored
during the 7-year duration of the MACHO microlensing project. A simple
autocorrelation calculation immediately showed that they all had
significant autocorrelation structure on time scales of several hundred
days and according to current theory such large structure must either
evidence systematic brightening of the small central source, whose light
travel time size is less than a light day, or must originate in distant 
outer structure illuminated by the central region. Because no theory for
central structure non-periodic oscillations with 100-day 
time scales is known to
us and reverberation between central and outer structure has been
reported on relevant time scales in Seyfert galaxies (Peterson et al 1993),
when scaled from Seyfert to quasar masses, we have presumed that the
effects we observe are the result of reverberation of the central continuum
brightness fluctuations on the outer luminous region
that has already been demonstrated by Elvis (2000) to originate the broad
blue-shifted emission lines due to an outflow wind. 

The data files that we analyzed are available at the Yale University website
www.astro.yale.edu/mgeha/MACHO. We immediately converted the brightness
records tabulated in magnitudes to linear units, and removed the outlying
data points exceeding 5 sigma. The brightness records available were
available from V and R color filters, but the V brightness records are
significantly more complete, and we have not thus far analyzed the R data.

Following our preliminary autocorrelation 
calculation we realized that approximately half of the quasars had
significantly more data points than the remainder.
To minimize problems associated with interpolation we restricted our
attention to the half which have the most data, each with
approximately 1000 nights. As we further analyzed the brightness data, we
found a variety of unexpected behaviors, such as semi-periodicity and
broad central autocorrelation peaks. Therefore we have chosen to report on
one particular quasar, MACHO 13.5962.237, which exemplifies the kinds of
behaviors encountered, to allow us to demonstrate the analysis techniques
employed. The redshift of the quasar is 0.17, and all results in this 
report are
expressed in time units measured by an observer in our local reference
frame.

\section{ Semi-Periodicity; An Underlying Source of Long Time Scale
Variability} 

Our initial procedure when confronting the MACHO quasar data was to compute
autocorrelation for each of the quasars, since we expected to see the
autocorrelation structure discovered in Q0957 by Thomson and Schild (1997).
However we noticed that for MACHO 13.5962.237 a much longer lag was seen to
the autocorrelation minimum at 700 days, and that the broad central
autocorrelation structure appears to show sub-structure as inflection points
for the 150-day lags expected. Inspection of the original brightness
record, Fig. 1a, showed that the largest brightness fluctuations were indeed
long-term changes that had the appearance of semi-periodic change.
In our Fig. 1a plot of the brightness data, the peaks of the semi-periodic
signature are at lags 200, 1600 and 2200 days. 
Correspondently, minima of the brightness curve are at days 900 and 2000.

Significantly, the central peak also showed inflection points for lags at
170, 280, and 450 days.

To look for the autocorrelation signal evidencing structure of the kind
seen in Q0957, we subtracted off the semiperiodicity. Because no model for
semi-periodic brightness fluctuations due, for example, to a relativistic
orbiting hot spot, and because simple inspection of the heavy solid curve
in Fig. 1a shows asymmetrical profiles, correction with a simple function
like $A sin t$ would not give a good representation and possibly leave
worrisome artifacts.

Therefore we adopted a simple procedure of removal by a kind of unsharp
masking. The structure we are looking for should have a time scale of
approximately 150 days and the observed quasi-periodicity autocorrelation
bottom is at 700 days. So we ran a 150-day boxcar smooth algorithm over 
the data to
create a smooth representation of the longest term trends. The smoothed
data is also shown in Fig. 1a as a heavy solid line. 
The smoothed data were then subtracted from
the original data to produce a brightness record dominated by
fluctuations on time scales of interest for reverberation, Fig. 1c.
Simple inspection of Fig 1c shows a pattern of sharp pulses with
brightness amplitudes of approximately 15 \% and durations of
approximately 50 days.

We next performed our autocorrelation analysis on the de-trended data, 
with the results shown in Fig 1d, where we clearly see a signficant
negative (anti-correlation) detection of 0.31 for lag 70 days. The profile
of this anti-correlation peak is convincingly symmetrical. The
anticorrelation peak is followed by several peaks of comparable width but
lower (0.1) positive correlation amplitude. 
Of the ensuing peaks shown, only the
first few are likely to be real, since the autocorrelation function
produces spurious peaks at multiples and at the sums of the shorter peak lags.

\section{ A Simple Model for Quasar Reverberation}

In previous study of the Q0957 +561 A,B gravitationally lensed quasar, we
found evidence for reverberation in the UV-optical continuum brightness 
curves and attributed it to reflections (or fluoresces) off of the outer
Elvis structure surfaces (Schild, 2005; SLR06). From this simple model, we
were able to determine the angle of the quasar rotation axis to the line of
sight. 

In Schild (2005) and in the absence of a full theory of brightness
fluctuations created from reverberation of a central disturbance, the
simple equations describing the sequence of pulse arrivals were presented
and solved only for the Q0957 quasar. In this section we show the full
solution in the form of a plot that shows how the reverberation pattern
varies with the angle of projection between the rotation axis and the plane
of the sky.

The equations for the time lags to the four brightness peaks expected for
the Elvis structure geometry with luminous regions above and below the
accretion disc plane have been given in Schild (2005).
For the purposes of this report, we adopt a
value of 13 degrees for the small internal quasar structure variable
$\epsilon$, since 13 degrees is the value determined for the two quasars for
which the reverberation has already been observed (Schild, Leiter, and
Robertson, 2006, 2008).

In figure 2 we show the plot of reverberation times for the four
reverberations seen in quasar brightness histories. We will see
in subsequent sections that the central quasar structure presumably causing the
starting pulse is apparently small, with a size parameter estimated from
the (half-)width of the central pulse of only 25 days. The reverberations have
typical time widths of 80 days and are separated by typical times of
100 days. 

The curves in Fig. 2 show behaviors that are qualitatively distinctive 
depending on the orientation of the quasar rotation axis to the plane of
the sky. For quasars seen pole-on (viewing angle 90 degrees), 
an initial brightening of the central
region (near to the inner edge of the accretion disc)
will be followed after some months by the brightening of the nearest ring,
and then by the farthest ring. For equator-on orientation, with the
rotation axis lying in the plane of the sky, the initial central pulse is
followed almost immediately by the brightening of the nearest surfaces and,
after a long lag, by brightening of the distant surfaces. For an
intermediate orientation, say 45 degrees, the four reverberations are well
separated in time and can be individually recognized easily, as has been
found for Q0957 where we have determined a viewing angle of 54 degrees (SLR06;
Schild 2005). 

The curves showing reverberation in Fig. 2 are shown having a width 
(fuzziness) approximately equal to the expected pulse width. So for a
quasar with a viewing angle of 60 degrees, shown as the horizontal line in
Fig. 2 (upper), we show in Fig. 2 (bottom) the pulse train expected.
An initial pulse of 25 day (half-)width is followed by
4 additional pulses having 80 day widths in the pattern shown.
The reverberation pulse amplitudes will be a complicated function of the
outflow wind geometry, and our Fig. 2 (lower) cartoon of the predicted
pulse train arbitrarily shows the pulses equal in brightness.

We have made an approximate solution for the orientation angle on the sky
for MACHO Quasar 13.5962.237. Although the equations for the reverberation
structure given in Schild (2005) are over-constraining, in the sense that 
there are four equations for 3 unknown parameters, we find that a solution
exists at 72 degrees for the four autocorrelation peaks identified from
the Fig 1d plot. The same autocorrelation estimate is found plotted to
larger scale in our final plot of comparison to simulated data. We do not
estimate a formal error bar because the errors are apparently systematic,
relating to small assymetries in the profiles of the pulses seen in
autocorrelation. However we suspect that from the agreement of solutions
for various assumptions about the determinations of the pulse centers that 
the true error of the determination is approximately 3 degrees.

\section{Determination of the Pulse Train for Reverberation Structure}

The positive autocorrelation peaks calculated from the brightness data are
of low amplitude, but are probably real. Since the original data defining a
single 70-day wide peak would contain on average 23 data points, each with
a 1-sigma error of 0.03 magnitude, an entire peak has a significance
exceeding 10 sigma. 

We have used a "poor man's autocorrelation" technique to reveal the nature
of the reverberations. In Figure 3 we have co-added the data segments
following the central structure peaks that we can easily identify in
the Fig. 1c brightness record. We have not included the wave trains
following several of the peaks,
especially in the cases where the peaks are double or ambiguous. In the
data segments in Fig. 3, we have placed the central brightness peak at 70
days, to see the average brightness preceding the peak in case there are
any precursors (which we do not find). In each panel of Fig. 3 (and of
Fig. 4) we show the individual wave trains following the initiating
pulse. For comparison, the mean wave train formed by averaging the 6
contributing wave trains is shown as a heavy solid curve. This mean curve
shows by construction the mean profile of the initial pulse and the ensuing
pattern of reverberations. The initial pulse has a mean average of 0.3
magnitudes, or 30 \% brightness peak, and a pulse (half-)width of 50 days.  
This is followed on time scales of approximately 100 days by secondary, or
reverberation, pulses, of comparable widths but lower, .10 mag or 10\%,
brightness increase. We do not know of any model that could generate an
error statistic for the mean waveform computed, because noise in this mean
waveform determination is physically caused by their overlapping.

Simple inspection of Fig. 1c also shows important events where the quasar
brightness $faded$, with the amplitudes and durations of the fading events
comparable or slightly smaller than the brightening structure. Such events
have never been predicted by theoretical work. The existence of such
events - and their characterization - must necessarily be ambiguous to some
extent because our methodology includes referring all fluctuations to the
long term mean quasar brightness. Nevertheless the fact that we find 
similar pulse train lags
for the fading and brightening events shows that they have 
some reality,
with future quantitative refinement of our analysis method still needed.

In Fig. 4 we show the procedure for averaging data segments for individual
fading event pulse trains analogous to Fig. 3. In particular, the peak fading
is again set at 70 days to allow inspection for a precursor. However in Fig. 4
all the data have been sign inverted to make the pulse trains look like our
brightening pulse trains in Fig. 3. It will be immediately recognized that
the Fig. 4 pulse trains have surprising similarity to the positive pulse
trains in Fig. 3. In particular, we find that the central structure has a
slightly lower mean amplitude but a similar pulse width. The later arriving
pulses have the same width as in Fig. 3, and are separated by comparable
lags. 

However an important difference is found between the pulse trains following
central brightening and fading events. We show in Fig. 5 a comparison
between the two mean pulse trains measured. In the comparison we find evidence
for an $inverted$ $structure$. This means that when the quasar brightens at a
reverberation site for some lag relative to a central brightening, the
quasar also $brightens$ at the same lag for a central fading. Such inverted
structure may be seen in Fig. 5 at lags of 82 and 115 days (shown by the
vertical lines in Fig. 5). The 82-day pulse in the two waveforms seem to
agree well in pulse width and amplitude (Fig. 5 bottom).
However for the 115-day lag, it is not obvious
if there is a real inverted structure or just a difference in waveform,
possibly caused by noise in the mean waveform determination.

In some ways, it may have been counter-productive to analyze a quasar
selected to show the full catalogue of brightness variability behaviors.
The resulting report has a remarkable list of unexpected behaviors needing
explanation and even though the data quality is excellent and does not
limit any of the interpretations, the complexity of the phenomena does.
Thus we present Fig. 6 with a view toward bringing together
several phenomena explored.

In Fig. 6 we show the brightness records of the mean wave trains determined
for the positive and negative spikes in comparison with the Fig 1d
autocorrelation calculation. The solid heavy line
represents the autocorrelation, and the thin line represents the mean
of the positive and negative pulse means. In other words, fig. 6 is a
comparison of the pulse trains estimated from autocorrelation with the
mean pulse train averaged from the many data segments. 
The purpose is to show the similarity of the
autocorrelation properties to the structured brightness trends.

In Fig. 6, the full width to the first 0 of the autocorrelation curve 
is at approximately 25 days. The zero crossing of the mean 
waveform curve occurs for a comparable lag. Thus the width of the 
initial brightness burst is measured to be about the same, 25 days, 
for both techniques. 

Summarizing our comparison of the autocorrelation estimates with our
attempts to identify the specific wave form of the brightness fluctuations,
we would say that we have found a mixed picture. While it seems likely that
a sharp 30\% brightness spike of 25-day width precedes a pulse train
extending over a year, the problem of overlapping of these wave trains
makes difficult their clear representation and
frustrates any attempt to estimate an error bar. The different kind of wave
train for pulses of fading brightness
may have related form. In particular, at the two lags where the brightening
and fading wave peaks have inverted structure, the autocorrelation estimate
has near-0 amplitude.

Nevertheless it is clear that some
kind of structure is evidenced in these pulse trains of year-long duration.
Because the standard black hole model of quasar structure does not include
any luminous regular structures with length scales longer than a light day,
the study of such repeating or quasi-repeating brightness features on
year long time scales is new information about the nature of luminous
quasar structure.

Although it might seem obvious that a brightening of quasar structure would
reverberate around the quasar as lagging brightenings, this is not
necessarily the case, and Gallagher and Everett (2007) have suggested a
mechanism that might produce negative reverberation.

\section{ The Quasar Fuelling Function}

In fig. 7 we show the amplitudes and lags of the quasar brightness pulses
determined as the cross-correlation of the mean pulse train profile
estimated in Section 5 with the
semi-periodicity corrected brightness data illustrated in Fig. 1c. With 
the interpretation that the pulse trains illustrated in
Figs. 3 and 4 represent the ordinary quasar brightness fluctuations 
resulting from its
fueling, the results in fig. 7 are effectively the quasar fueling function.
Recall that the curve in Fig. 7 has been set againsst a zero mean.

The results in Fig. 7 seem to show that the quasar fueling is a
reasonably stochastic process with a train of pulses having similar
brightness amplitude and duration during the 7-year period of
observations. 

\section{Noise Simulation}

Because of the complexities of the brightness phenomena that we are
discussing, and particularly because no model of these effects is yet
available, it is difficult to imagine a thorough simulation of the
results obtained in the context of random noise. However, we have
undertaken the two simplest kinds of simulations to make some approximation
to the effects of random noise.

Our first simulation was made by simply scrambling the data points randomly
onto  the observed observation dates, to produce a noise data set related
to the set we have been analyzing. This produced very noisy plots of
autocorrelation which reached a peak anti-correlation value of -0.1
(instead of our comparable observed value of -0.45) and a significantly
different appearance of the autocorrelation peak. The peak in the
simulation was significantly narrower, at about 10 days width, which is the
effective resolution of our data. 10 realizations of this randomizing of
the data points produced similar results that did not look like our real
data products.

A second simulation was made by following our next procedure and producing
a boxcar smoothed version of our data and subtracting it from our
simulated data set. This produced autocorrelation and mean wavelet
properties about as above but with lower amplitude. Nevertheless the
autocorrelation peaks had much sharper structure than our real data, and in
10 realizations gave qualitatively different results than our real data.

In Fig. 8 we show a typical plot of the noise simulation for the quasar
data with red noise suppressed exactly as was done for our real data. The
solid curve is the autocorrelation computed for real data. It may be seen
that the two are qualitatively different, with the quasar data showing 
quasar brightness fluctuations on time scales much slower than the
effective sampling, whereas the noise data show twinkling at lower
amplitude and higher frequency.

\section{ Conclusions and Discussion}

We have found that the MACHO program observations of quasar brightness are
an excellent resource for the study of brightness fluctuations, because the
fields observed are circumpolar and do not suffer from seasonal dropouts
and produce results independent of procedures of interpolation. From our
autocorrelation analysis of 29 of the best studied cases, we have found
that fluctuations seem to have interesting characteristics on two major time
scales. 

On the longest time scales, of order of 2 years, the fluctuations seem to
have a quasi-periodic character with amplitudes of order 30\% brightness
change. They are not well represented by sine (t/500 days) but nevertheless
seem to have a quasi-regular structure. We have analyzed one of the quasars
showing this structure and investigated further the structure seen on much
shorter time scales. Although it may be premature to call the large overall
brightness structure semi-periodic on the basis of the data for MACHO
13.5962.237, similar structure has already been seen in the much longer
brightness records for the Q0957 quasar (Pelt et al, 1998, Figs. 1 and 2). 
Though seen at
a larger redshift, 3 such peaks with comparable 0.3 magnitude amplitude and
reasonably uniform cadence have been seen.

After removing the quasi-periodic overall structure, we have found that the
autocorrelation properties evidence complex and reasonably regular
structure having the character of a sharp central pulse followed by
broader secondary pulses of lower brightness. The pulses are surprisingly
similar for these rapid events whether the quasar is brightening or fading.

It is worth stating even though not proved herein, that all of the 30
quasars for which we have analyzed brightness data evidence structure on
50-day and separately multi-year time scales. Thus large scale structure 
with associated size scales of a light year appear to be a universal
property of quasars. Because the tentatively observed reverberations
have been associated with the previously discussed region where the
quasar-defining broad high-ionization blue-shifted emission lines
originate, it appears that only a fraction of the dominant UV-optical 
emission of quasars comes from the central region understood
theoretically. 

Our results suggest a technique for discovering quasars from large area
photometric surveys presently being planned (Pann-Stars, LSST). Calculation
of auto-correlation of photometric brightness curves will reveal a signal 
with anti-correlation bottom of at least -.2 for quasars. It remains to
undertake simulations of data sets at various noise levels and
observational epochs to establish the limits of effectiveness of this
quasar discovery technique.

Such photometric monitoring has the prospect of not only discovering
quasars, but also of directly determining masses. Establishment of the
fundamental plane for MECO quasars by Leiter and Schild (in preparation)
seems to show that measurement of the bolometric brightness and a
reverberation-measured
size parameter related to the distance of the quasar's light cylinder, or to
the size of the inner edge of the accretion disc, allows the mass of the
central object to be determined.

\section{Acknowledgements}

We thank Tesvi Mazeh and Victor Vakulik for useful conversations about the
properties of the autocorrelation function.

\newpage
\begin{figure}
\begin{center}
\plotone{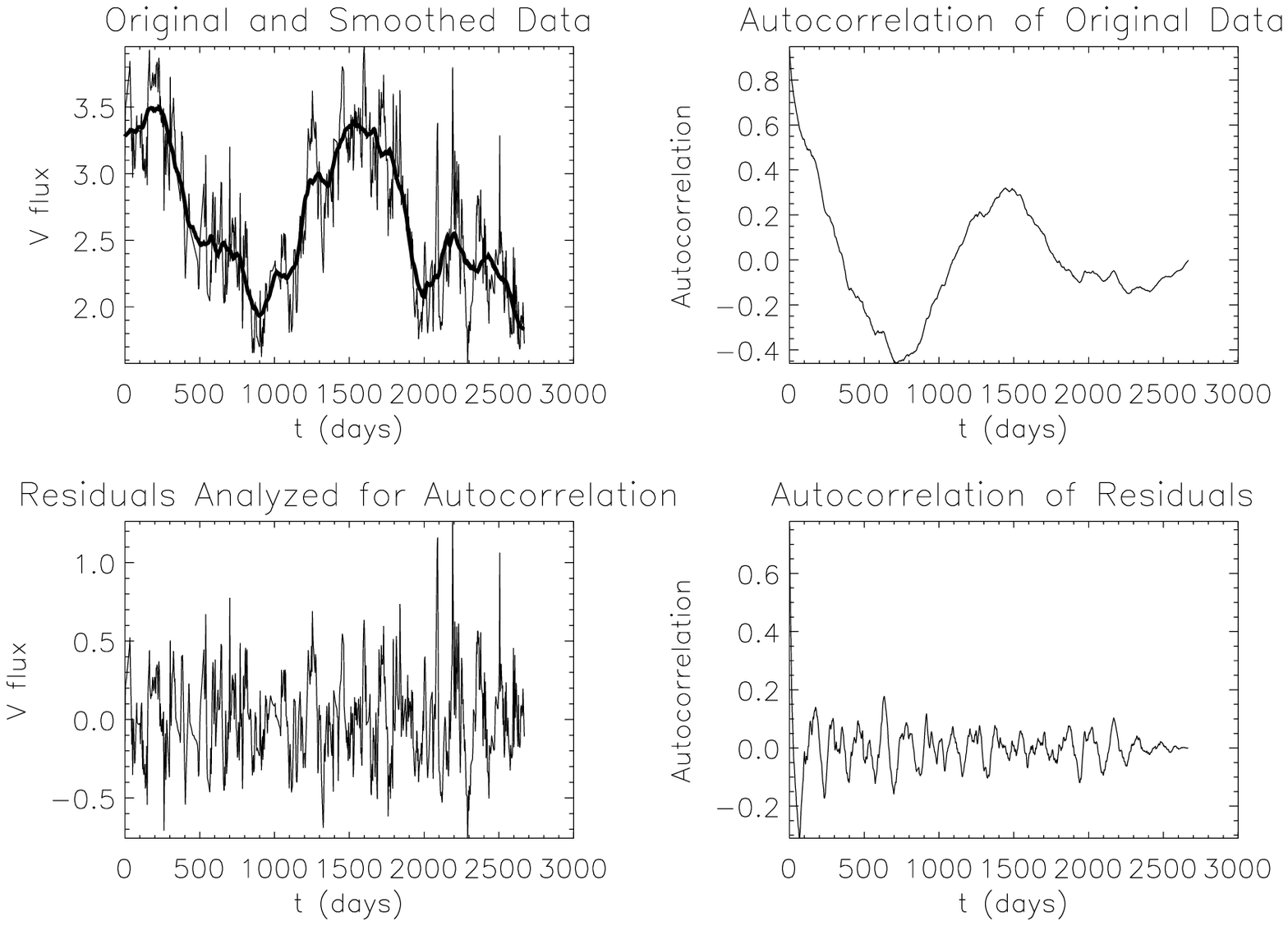}
\caption{(a) Plot of the original data for M13.5962.237 with a
150-day boxcar smoothed version of the data overplotted as a double-weight
line. (b) A simple autocorrelation calculation of the original
data shows a significant autocorrelation bottom at 710 days. (c)
When the boxcar smoothed curve is subtracted from the original data,
effectively filtering out the lowest frequencies, a plot of the
high-frequency component dominated by reverberations is obtained.
(d) The auto-correlation of the high-frequency component now
shows a significant anti-correlation bottom at only 80 days, and a strong
narrow central autocorrelation feature.}
\label{fig. 1}
\end{center}
\end{figure}

\newpage
\begin{figure}
\begin{center}
\plotone{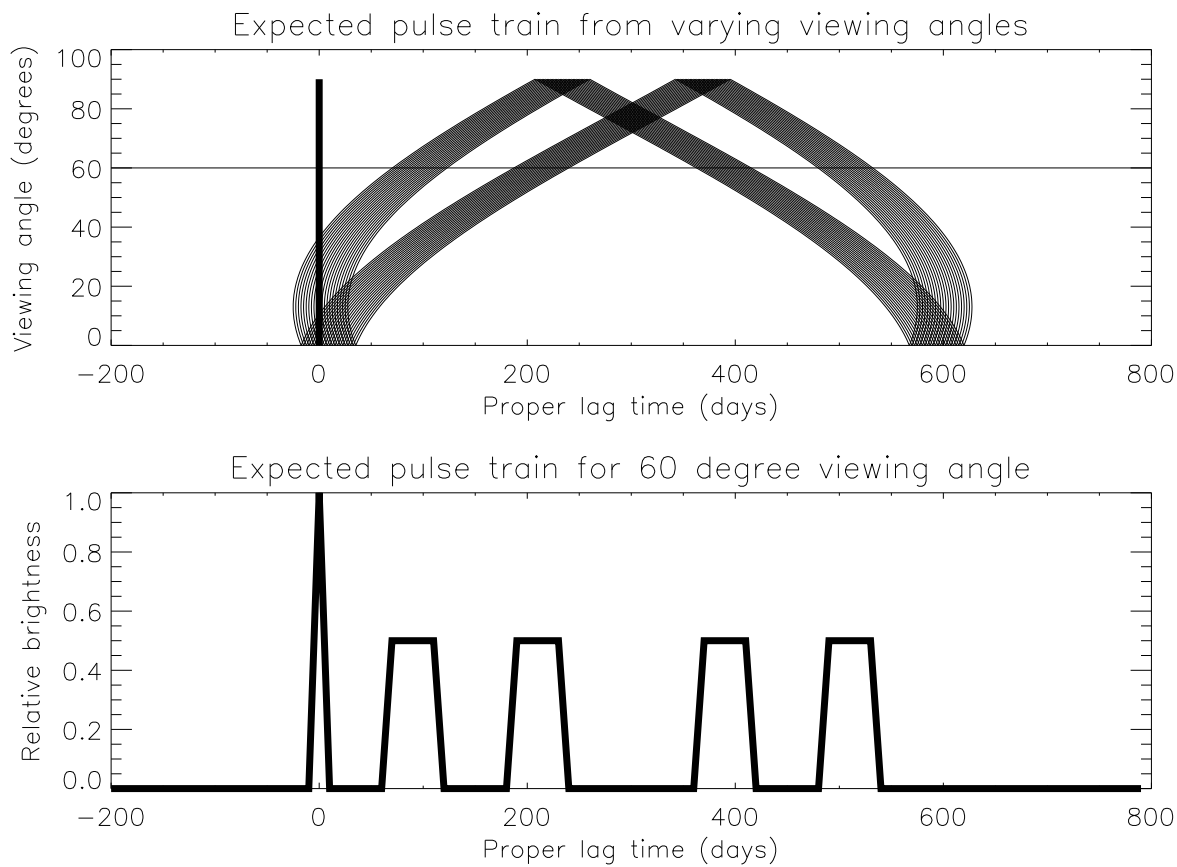}
\caption{(Upper) A plot showing the order of reverberation pulses expected 
as a function of the viewing angle (between the plane of the sky and the
axis of rotation) of the quasar. The sharper initiating pulse is shown as a
triple-weight line at t = 0, and subsequent broader fainter pulses of
reverberation follow at the times shown. For quasars viewed along 
their equators
the viewing angle is 0 and the initial pulse is broader because it occurs at
the same time as the reverberations from the nearest two surfaces. A
horizontal line at viewing angle 60 degrees shows how the plot would be
read to show the pattern of reverberation features expected.
(Lower) We show the pulse train expected for a quasar inclined at 60
degrees, according to the structures computed in the upper panel. The
amplitudes of the secondary pulses will vary as a function of viewing
angle, and are shown all the same in the cartoon for simplicity} 
\label{fig. 2}
\end{center}
\end{figure}

\newpage
\begin{figure}
\begin{center}
\plotone{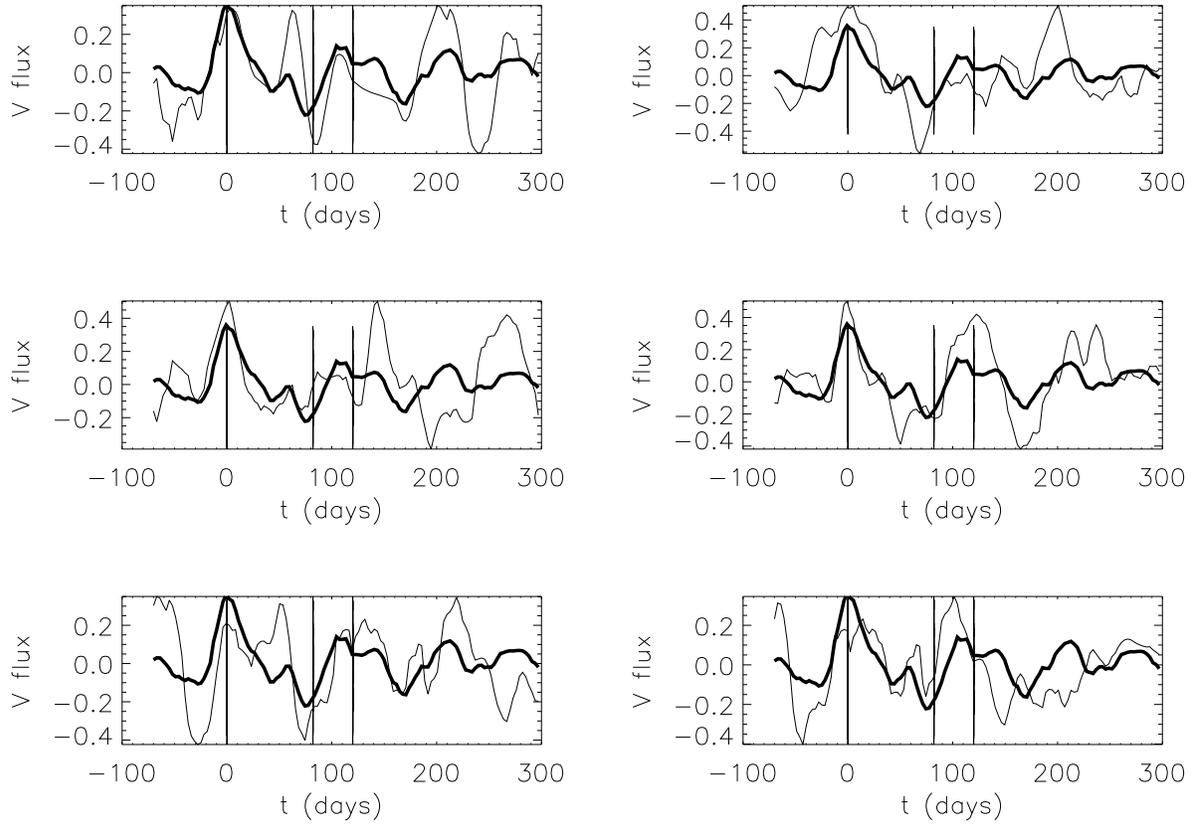}
\caption{Data segments co-added to define the shape of the waveform from
reverberation. Six data segments are co-added to form the mean wave form
shown as the solid heavy curve in all panels. }
\label{fig. 3}
\end{center}
\end{figure}

\newpage
\begin{figure}
\begin{center}
\plotone{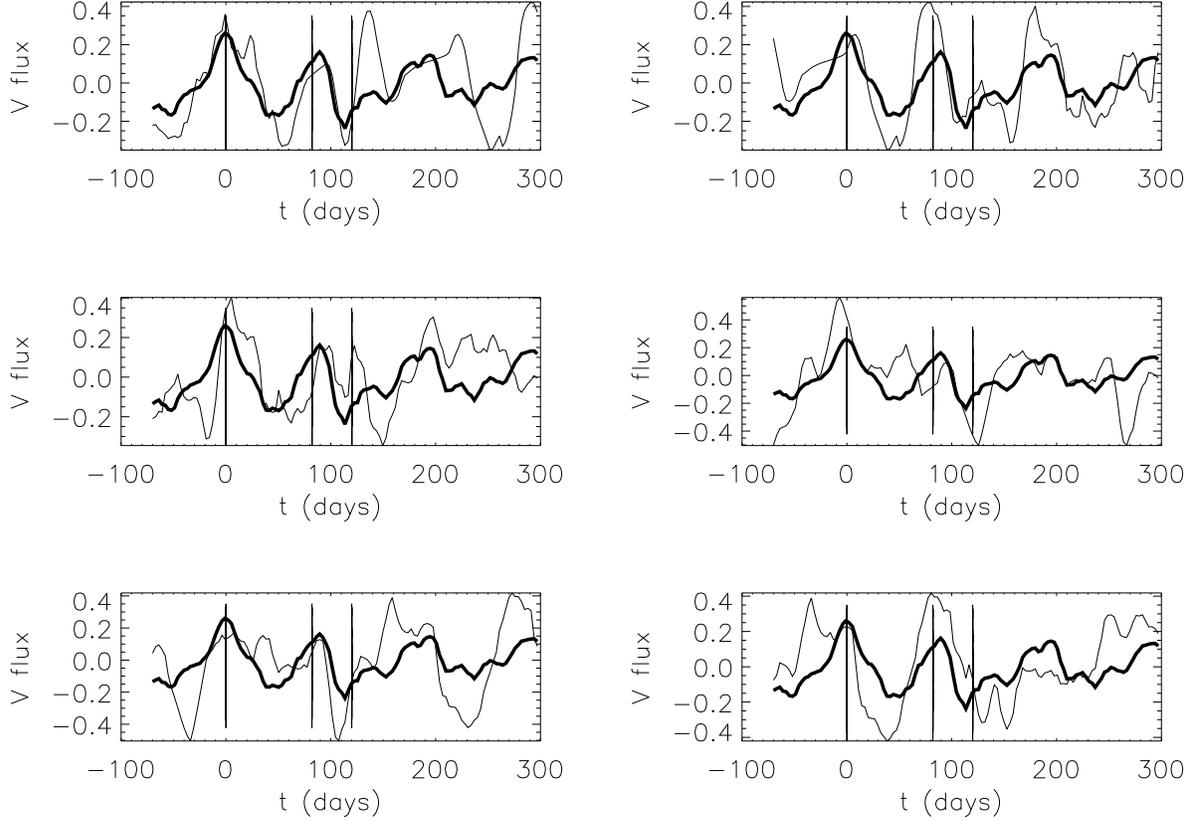}
\caption{Six data segments co-added as in fig. 3 but for the negative-going
reverberation pulse shapes, with the mean waveform shown as a solid heavy
curve. The large diversity of brightness curves results not from
observational noise, but from overlapping of the many pulse trains. The
data have been plotted inverse to the observations, meaning that the fading
pulses have been shown as positive, to allow simple comparison to the
positive pulse waveform. Vertical lines show the lags where inverted
structure may be present between the brightening and fading mean waveforms} 
\label{fig. 4}
\end{center}
\end{figure}

\newpage
\begin{figure}
\begin{center}
\plotone{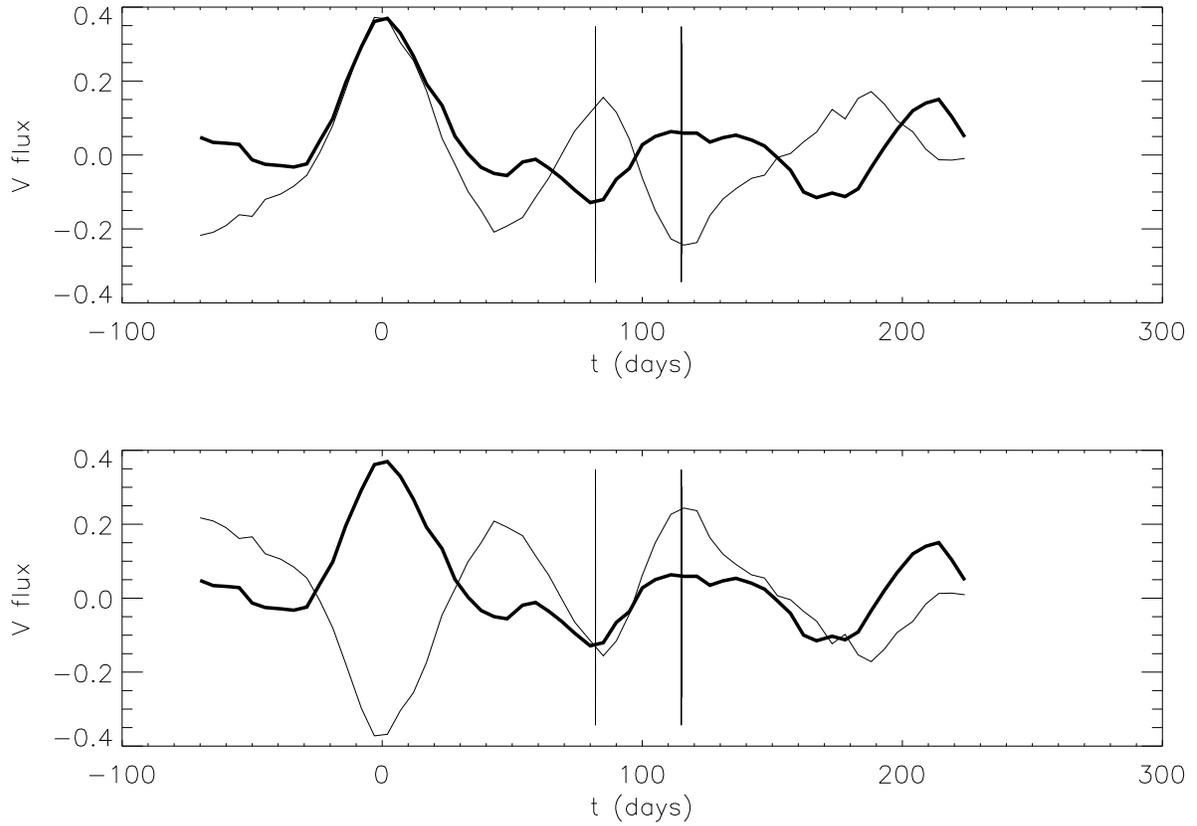}
\caption{ (Top)A comparison of the positive (solid heavy curve) and negative
  (light curve) mean wavelets,
  demonstrating the inverted structure for lags 80 and 115 days (vertical
  lines). This quasar was selected as an excellent example of such inverted
  structure, found in only 20 \% of the the quasars. (Bottom) The wave
  forms are shown again, but with the correct sense of brightening and
  fading pulses preserved.}
\label{fig. 5}
\end{center}
\end{figure}

\newpage
\begin{figure}
\begin{center}
\plotone{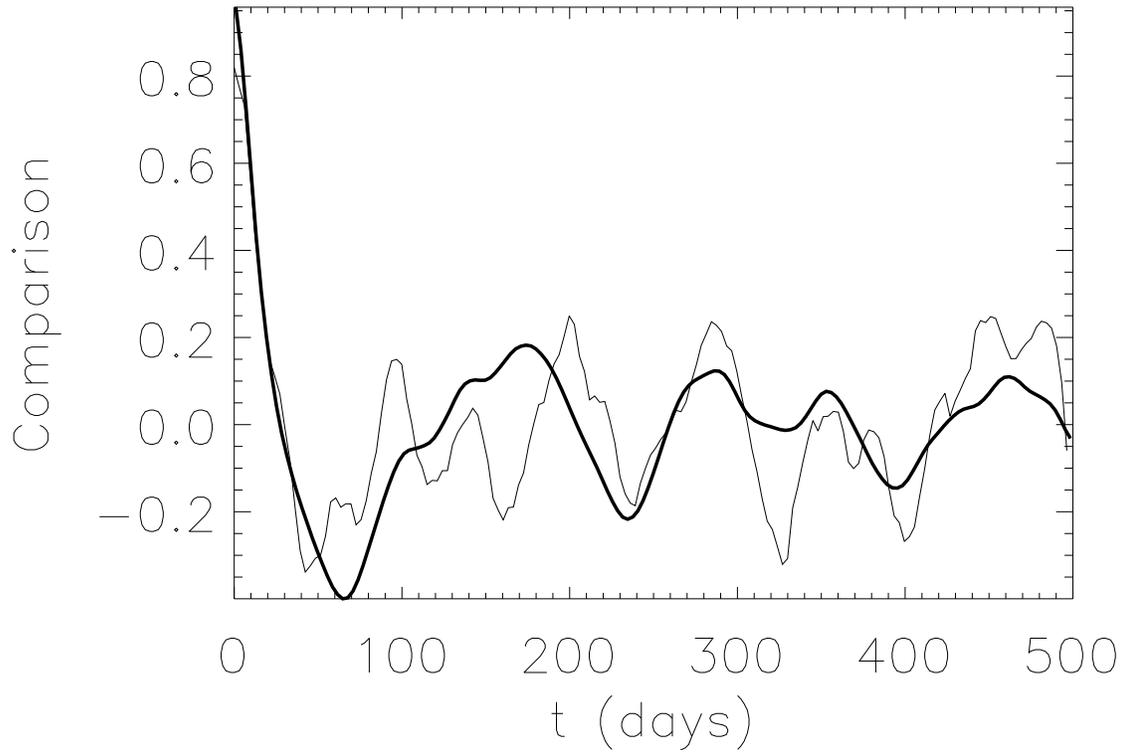}
\caption{The mean reverberation waveform from averaging the positive-going 
and negative-going wave forms, compared with
 the autocorrelation function for the high-frequency data.}
\label{fig. 6}
\end{center}
\end{figure}

\newpage
\begin{figure}
\begin{center}
\plotone{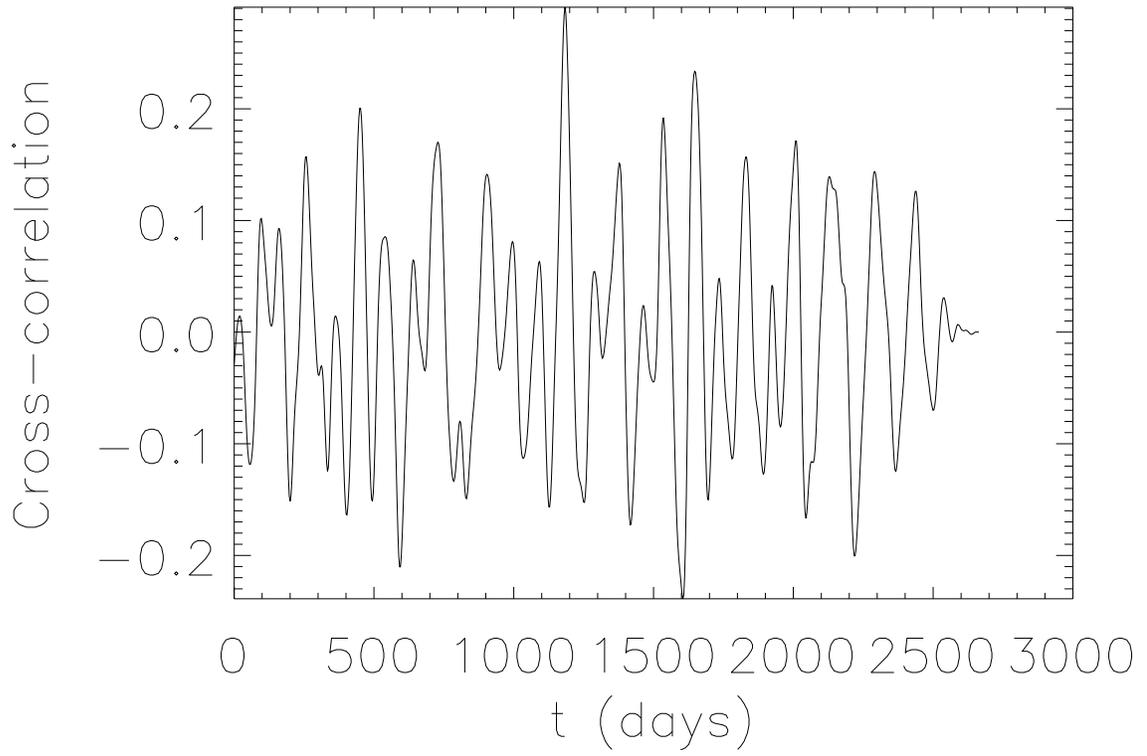}
\caption{Determination of the fueling function as the autocorrelation of
  the mean reverberation waveform against the low-pass filtered brightness
  record. }
\label{fig. 7}
\end{center}
\end{figure}

\newpage
\begin{figure}
\begin{center}
\plotone{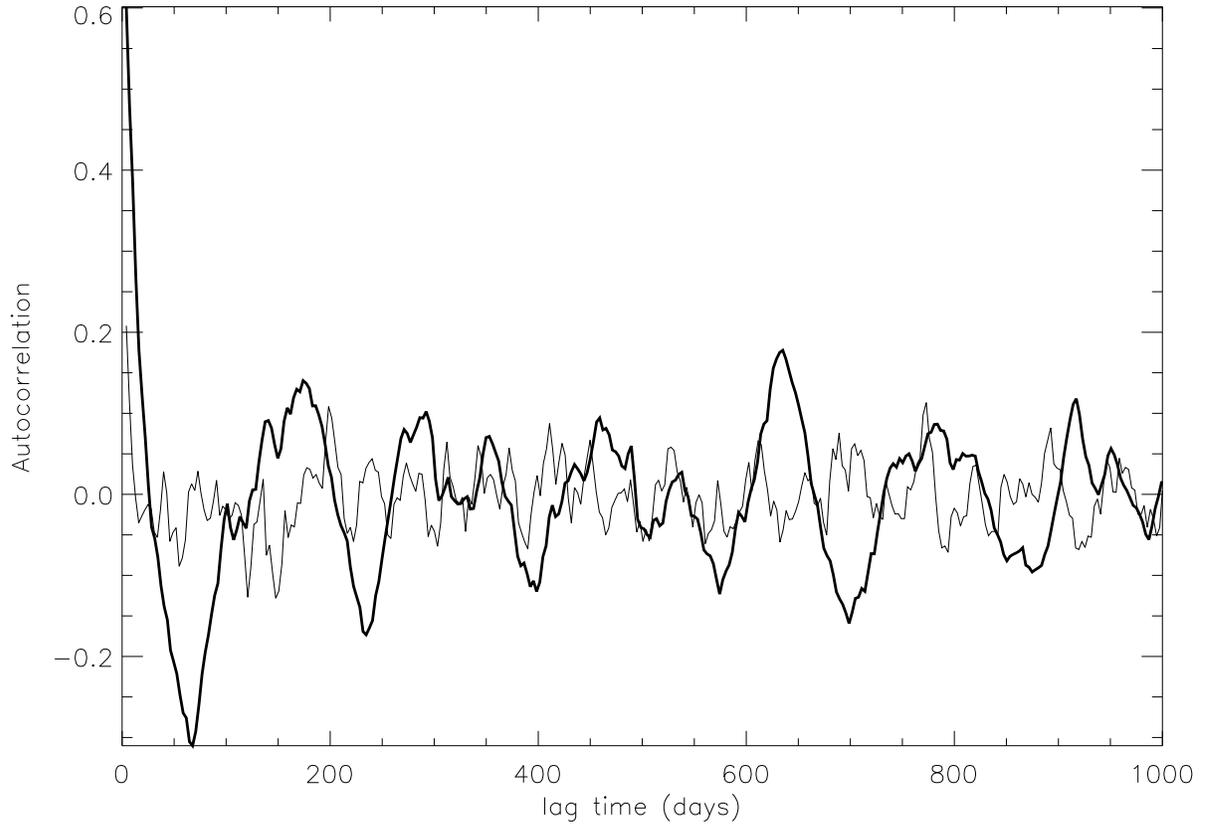}
\caption{A noise simulation for the autocorrelation function, from data
  created by scrambling the lags in the table of observations. The
  simulated noise calculation has been processed with the same noise
  averaging and smoothing before autocorrelation. Compared to the
  autocorrelation calculated for the real data, plotted as the heavy solid
  line, the simulated data show qualitatively lower amplitude of
  autocorrelation and structure on much shorter time scales. }
\label{fig. 8}
\end{center}
\end{figure}

\end{document}